\documentclass[12pt,a4paper]{article}

\usepackage{graphicx}\usepackage{amsmath}

\usepackage{subfigure}

\begin{document}

\begin{center}{\bf \Large The Galam Model of Minority Opinion Spreading and the Marriage Gap}\\[5mm]

{\large  Krzysztof Ku{\l}akowski$^1$ and Maria Nawojczyk$^2$\\[3mm]

\em {$^1$ Faculty of Physics and Applied Computer Science, $^2$ Faculty of Applied Social Sciences, AGH University of Science and Technology, al. Mickiewicza 30, 30-059 Krak\'ow, Poland\\

E-mail: kulakowski@novell.ftj.agh.edu.pl\\\today}}

\end{center}

\begin{abstract}

In 2002, Serge Galam designed a model of a minority opinion spreading. The effect is expected to lead 
a conservative minority to prevail if the issue is discussed long enough. Here we analyze the marriage 
gap, i.e. the difference in voting for Bush and Kerry in 2004 between married and unmarried people.
It seems possible to interpret this marriage gap in terms of the Galam model.

\end{abstract}

\section{Introduction}

Whatever the definition of a complex system would be, the complexity of a social system cannot be denied.
It excess not only technical possibilities of our research tools, but also our imagination. To describe 
a social systems, the sociophysicists try to apply ideas of statistical physics: interaction, phase 
transition, equilibrium, self-organized criticality, noise. The results are sometimes very inspiring,
but scepticism can hardly be defeated. Some reasons of this state of art are obvious. Social systems cannot 
be isolated, therefore experimental results are not reproducible and the idea of a measurement is undermined.
For the discussion on application of experiment in social research see Ref. \cite{bab}.
Moreover, societies are always finite, always biased; no symmetry is preserved, any equilibrium is not 
possible. Elements of social sets differ one from another and they strongly interact one with another. 
Even if a good model of the society appears, some people apply it for their purposes and forthwith it ceases
its validity. Despite of all that, to investigate social systems is more than necessary. Moreover, we share 
the belief that the quantitative description of social systems, although risky, is desired. As this opinion
seems to be common between the sociophysicists, strong assumptions are routinely applied in order to capture 
social processes in terms of most simplified models. 

Here we refer to the Galam model of the dynamics of opinion formation in a society \cite{g1,g2}. Main goal of 
these works was to demonstrate a social mechanism, what leads some kind of minority opinion to prevail. 
This prevailing opinion "has to be consistent with some larger collective social paradigm" of a given society.
Galam gives examples: the failure of French government to reform the academic system in 2000, Irish "No" to the 
Nice European Treaty, fear against social reforms \cite{g1} or the anti-American rumor on September 11' 
attack, spreading in France \cite{g2}. In all cases, some prejudice widespread in the society is expected to support
the minority opinion. The mechanism of shaping the public opinion by rumors connetced with stereotypes is
very well described in social psychology \cite{alpo}.
Mathematically, the idea is that people adopt their opinion to the majority in small random 
groups they gather in. The bias mechanism that works for the initially minority opinion is - according to Galam - 
that in the case of a tie, this opinion is adopted by the group. As a tie is possible only in groups with even number 
of people, these groups play the main role in the story. For a good review of the mathematical and computational 
approaches to the Galam model see Ref. \cite{totes}. 

Direct results of the model depend on its mathematical realization. In particular, one can ask about the critical 
percentage $p_c$ of the minority necessary to trigger the process. If the initial minority size is less than $p_c$, 
the opposition dies out; if it is larger, the minority opinion prevails. The consequence of the bias is that $p_c<0.5$.
The value of $p_c$ for groups of equiprobable sizes 2, 3 and 4 was found to be 0.15 \cite{g2}. In our opinion, 
more important is to ask whether the assumed mechanism does exist. Does the human tendency toward conformity 
as well as cognitive misery \cite{aro} within group life impose a conservative attitude?

The aim of the present note is to provide an example which could verify the presence of the mechanism described by
Galam. First we have to look for social groups with even number of members; an example that suggests itself is 
the marriages. Looking for an opinion which could be considered as conservative, as a kind of a prejudice against
changes, we found the data on the Bush-Kerry race in 2004 presidential elections \cite{abc}. In this poll, Bush
won by 20 points among married men and by 16 points among married women. Conversely, single men favored Kerry
by 65:32, and so did unmarried women (59:38). The effect has been referreed to as 'marriage gap', although this 
term is used also in other cases when being married apparently correlates with some statistical feature. Similarly, 
a 'gender gap' has been noticed during earlier presidential votes in USA.

In the next section we present the Galam model limited to marriages. On the contrary to the original formulation
given in \cite{g1,g2} we are going to assume that these groups of two persons are constant in time; this assumption 
seems realistic at least in the timescale of an electoral campaign. We will show that in this case the model 
predictions can be obtained in a one-line calculation. Last section is devoted to discussion.

\section{Opinions of marriages}

In the Galam model, people are expected to modify their opinions by accepting the majority in a group, where they happen 
to meet. Closing the model with that, we get a simple result that the majority always wins, and $p_c$ = 0.5. As remarked
in the previous section, the bias comes from the cases where equal numbers of group members vote pro and contra - the set
of opinions is limited to these two. Then, an opinion prevails that is consistent with some widespread attitude of the 
society. As indicated in \cite{totes}, the original model \cite{g1,g2} neglects spatial correlations; this limitation was 
removed in subsequent papers where computer simulations were applied to the Galam model (\cite{sta,totes}). In these 
approches, opinions are carried by people from one group to another and spread in a stochastic way.

Our reasoning is as follows: let us assume that the mechanism described by Galam exists. Then it should be active not 
only in random contacts, but also in small permanent groups. We can capture it statistically if these groups are well 
defined; also, they should consist of even number of people. The marriages seem to be a good example; if the effect cannot
be seen with the marriages, it is likely that it does not appear at all.

We also need a poll on an issue, where the answers can be qualified as conservative or not. However, the term 
'conservative' itself is ambiguous and its discussion could fill volumes. We must base on its common understanding,
supported by an additional association with a widespread prejudice. We hope that our example - the Bush-Kerry race in 2004 - 
is not far from the intention of Galam's texts.

An application of the idea to permanent groups of two persons leads to the simple argument that the statistics will be 
modified as follows. Let us treat the attitude of unmarried persons as unbiased. We need just the probabilities
that an unbiased (=unmarried) woman or man votes for Bush and Kerry. According to Ref. \cite{abc}, these probabilities 
are: $m_B$=0.32 (man-Bush), $m_K$=0.65 (man-Kerry), $w_B$=0.38 (woman-Bush) and $w_K$=0.59 (woman-Kerry). These 
probabilities are not normalized, but $m_K+m_B=w_K+w_B$; then the ratio of votes for Kerry and Bush is not changed.
This ratio for the unmarried is Bush : Kerry = $(w_B+m_B)$ : $(w_K+m_K)$. The same ratio is expected for the 
married people before they start to discuss the issue. In the discussions, the probability of two pro-Bush's (wife and
husband) is $w_Bm_B$, and they preserve their opinion; the same with two pro-Kerry's, with the probability $w_Km_K$. The 
idea of the bias is that the couples with different opinions will vote for Bush. Denoting the percentage of married 
people by $b$, the final output is

\begin{equation}
(1-b)(m_B+w_B)+2b(m_Bw_B+m_Bw_K+m_Kw_B) : (1-b)(m_K+w_K)+2bm_Kw_K
\end{equation}
In this calculation, correlations between husband and wife are neglected in the initial state, and perhaps overevaluated
in the final state of the time evolution. The amount $1-b$ of unmarried people in USA is close to 0.41 \cite{stat}. Then, the 
obtained formula, when normalized, gives the ratio 49.5:50.4, what is not far from the final tie. However, the amount of 
married voters in the poll of voters is somewhat larger, 0.63 \cite{cnn}. In this case Bush prevails 50.5:49.5.

To find the proportions for the married people, we do not need to rely on the value of $b$. The marriages vote 
for the same candidate, then the probabilities of a married person to vote for Bush or Kerry is $(w_Bm_B+m_Bw_K+w_Bm_K):m_Kw_K$. 
The result is 56:38, what nicely fits the data of Ref. \cite{abc}: 56:40 for the married women and 59:39 for the married
men.

\section{Discussion}

All what was told at the beginning of this paper warns us not to take these numbers too seriously. They should be treated
not as a proof that the investigated mechanism exists, but rather as a statement that this mechanism is not excluded by the
presented data. The argumentation can be attacked at numerous points. 

For example, every married person is convinced that an accordance of a couple can hardly be assumed as a starting point. 
However, the ability of cohesive group to influence its members and the consequences of this phenomenon had been for a 
long time an interesting issue of investigation in micro-sociology \cite{m1,m2}. Is the marriage conservative {\it ex 
definitione} ? The research done by sociologists and demographers show that alternative scenarios of married life are 
more and more attractive for various groups of people for different reasons \cite{m3}. Also, one can wonder why the 
marriage gap appeared only recently?  Is it an indication of the emergence of post-family family using Becks term 
\cite{m4} as one of the features of political differences among Americans voters? Further, to what extent is the 
Republican Party conservative? Discussion of these questions would lead us far beyond the frames of this work. Our results indicate, that it makes sense 
to ask voters if their families share their opinions. Such data could provide a better insight into the mechanisms of our 
political decisions.

To conclude, we share the opinion expressed by Hans Bethe that it is necessary to look for a verification of mathematical 
theory. The models should be constructed as to enable such a verification. As we could demonstrate, the 
mechanism proposed by Galam fulfills this criterion. \\

{\bf Acknowledgements.}   The work was partially supported by COST Action P-10 "Physics of Risk".


\begin{thebibliography}{00}

\bibitem{bab} E. Babbie: \textit{The Basics of Social Research}, (Wadsworth Publishing Co., Belmont 1999)
\bibitem{g1} S. Galam: Eur. Phys. J. B \textbf{25}, 403 (2002)
\bibitem{g2} S. Galam: Physica A \textbf{320}, 571 (2003)
\bibitem{alpo} G. W. Allport, L. Postman: \textit{The Psychology of Rumor}, (Holt, Rinehart and Winston, New York 1947)
\bibitem{totes} C. J. Tessone, R. Toral, P. Amengual, H. S. Wio, M. San Miguel: Eur. Phys. J. B \textbf{39}, 535 (2004)
\bibitem{aro} E. Aronson: \textit{The Social Animal}, (W. H. Freeman and Co., New York 1992)
\bibitem{abc} G. Langer: ABC News, October 24, 2004 (http://abcnews.go.com/Politics/print?id=193178)
\bibitem{sta} D. Stauffer: Int. J. Mod. Phys. C \textbf{13}, 975 (2002)
\bibitem{stat} http://www.census.gov/
\bibitem{cnn} http://www.cnn.com/ELECTION/2004/pages/results/states/US/P/00/epolls.0.html
\bibitem{m1} R. S. Wyer: J. of Personality and Social Psychology \textbf{4},  21 (1966)
\bibitem{m2} G. C. Homans: \textit{Social Behavior. Its Elementary Forms}, (Harcourt-Brace-Jovanovich, New York 1974)
\bibitem{m3} K. Davis: Bull. Amer. Acad. of Arts and Sciences \textbf{36}, 15 (1983)
\bibitem{m4} U. Beck, E. Beck-Gernscheim: \textit{Individualization, Institutionalized Individualism and its Social and 
Political Consequences}, (Sage, London 2002)

\end{thebibliography}
\end{document}